\documentclass[oneside,a4paper,reqno]{amsart}
\usepackage{amssymb}

\newtheorem{theorem}{Theorem}
\newtheorem{lemma}[theorem]{Lemma}
\theoremstyle{remark}
\newtheorem*{remark}{Remark}
\DeclareMathOperator{\Tr}{Tr}

\DeclareMathOperator{\Det}{Det}

\begin{document}
\begin{titlepage}
\begin{center}
\bfseries  SYMMETRIC INFORMATIONALLY COMPLETE MEASUREMENTS OF ARBITRARY
RANK
\end{center}
\vspace{1 cm}
\begin{center} D M APPLEBY
\end{center}
\begin{center} Department of Physics, Queen Mary
University of London,  Mile End Rd, London E1 4NS,
UK
 \end{center}
\vspace{0.5 cm}
\begin{center}
  (E-mail:  D.M.Appleby@qmul.ac.uk)
\end{center}
\vspace{0.75 cm}
\vspace{1.25 cm}
\begin{center}
\vspace{0.35 cm}
\parbox{12 cm }{There has been much interest in so-called SIC-POVMs: 
rank $1$ symmetric informationally complete positive operator valued
measures.  In this paper we discuss the larger class of
POVMs which are symmetric and informationally complete but not
necessarily rank
$1$.  This  class of POVMs is of some independent
interest.  In particular it includes a POVM which is closely related to
the discrete Wigner function.  However, it is interesting mainly
because of the light it casts on the problem of constructing rank 1
symmetric informationally complete POVMs.  In this connection we derive
an extremal condition alternative to the one derived by Renes \emph{et
al}.    }
\end{center}
\end{titlepage}
\section{Introduction}
There has been much interest in rank $1$ symmetric, informationally
complete positive operator valued
measures~\cite{Renes,
FuchsA,WoottersC,BengA,GrasslA,BengB,
ApplebyB,GrasslB,KlappA,KlappB,Scott,Flammia,Kim}:   
SIC-POVMs, as they
are often called.  In
$d$-dimensional Hilbert space these are systems of $d^2$ operators
$\hat{E}_r = (1/d)\hat{P}_r$ such that each $\hat{P}_r$ is a rank $1$
projector and
\begin{equation}
\Tr(\hat{P}_r\hat{P}_s) = \frac{1}{(d+1)}(1+\delta_{r
s})
\end{equation}
for all $r,s$.  In that case it can be shown that $\sum_{r=1}^{d^2}
\hat{E}_r=1$, so the operators $\hat{E}_r$ constitute a POVM.  Moreover
the POVM is informationally
complete~\cite{Prugo,Schroeck,BuschA,BuschB,Caves,FuchsB,dAriano}
(meaning that an
arbitrary density matrix
$\hat{\rho}$ is completely specified by the probabilities
$\Tr(\hat{E}_r \hat{\rho})$).  
POVMs of this kind have been
constructed~\cite{Renes,GrasslA,ApplebyB,GrasslB,Hoggar,Zauner}
(analytically and/or numerically) for every dimension $d\le 45$.  It is
still an open question whether they exist in dimensions $>45$.  

The purpose of this paper is to discuss POVMs which are still symmetric,
in the sense that
\begin{equation}
\Tr(\hat{E}_r \hat{E}_s) = \alpha + \beta \delta_{rs}
\end{equation}
for fixed numbers $\alpha, \beta$, and informationally complete, but which
are not assumed to be rank $1$.  We will refer to such POVMs as SI-POVMs
(``S'' for ``symmetric'', ``I'' for ``informationally complete''). 
SI-POVMs which are also
 rank
$1$ we will refer to as SI(1)-POVMs (so an SI(1)-POVM is what in the
literature is often called a SIC-POVM).

SI-POVMs are of some independent interest.  In
particular, we will show in Section~\ref{sec:WigPOVM} that the discrete
Wigner function is closely related to a POVM of this type.  However, our
main reason for studying them is to  gain additional insight into the
problem of constructing SI(1)-POVMs.  To that end we derive an extremal
condition alternative to the one used by Renes \emph{et al} in their
numerical work.  

The plan of the paper is as follows.  In Section~\ref{sec:BlochBody} we
discuss some geometrical features of quantum state space which will be
needed in the sequel.  In Section~\ref{sec:BlochAndToography} we relate
this discussion to the problem of devising a tomographical procedure
which is, in some suitably defined sense, optimal.  In
Section~\ref{sec:SymmGeneral} we prove a theorem characterising the
structure of an arbitrary SI-POVM.  In
Section~\ref{sec:WHCovariantSIPOVMs} we specialise to the case of SI-POVMs
covariant under the Weyl-Heisenberg group (or generalized Pauli group as it
is often called).  We show that such POVMs have a very simple
representation in terms of the Weyl-Heisenberg displacement operators. In
Section~\ref{sec:SI1Construct} we turn to the problem of constructing
SI(1)-POVMs, and derive an extremal condition alternative to the one
derived by Renes \emph{et al}~\cite{Renes}.  Finally, in
Section~\ref{sec:WigPOVM} we construct an SI-POVM which is closely
related to the discrete  Wigner function.

\section{The Bloch Body}
\label{sec:BlochBody}
Let $\mathcal{H}$ be a $d$ dimensional Hilbert space, and let
$\mathcal{D}$ be the space of density matrices defined on $\mathcal{H}$. 
If $d=2$ it is well known that $\mathcal{D}$ can be identified with the
Bloch sphere.  To be specific:  let $\mathcal{B}$ be the unit ball in 
$\mathbb{R}^{3}$ (\emph{i.e.}\ the set of vectors $\in
\mathbb{R}^3$ having length $\le
1$).  Then
a
$2\times 2$  complex matrix
$\hat{\rho}$ is a density matrix if and only if it can be written in the
form
\begin{equation}
\hat{\rho}=\frac{1}{2} \bigl(1+\mathbf{b}.\hat{ \boldsymbol{\sigma}} 
\bigr)
\label{eq:Bloch2D}
\end{equation}
where $\mathbf{b}\in \mathcal{B}$ (the Bloch ball) and
$\hat{ \sigma_1},\hat{ \sigma_2},\hat{ \sigma_3}$ are the Pauli
matrices. 

With the appropriate modifications this construction can be generalized to
higher
dimensions~\cite{BengA,BengB,Harriman,Mahler,Jakobczyk,
Kim1,Byrd,Schirmer,Kim2,Dietz}.  
 Let $\mathrm{su}(d)$ be
the $d^2-1$ dimensional real vector space  consisting of all trace zero
Hermitian $d\times d$ complex  
matrices\footnote{$\mathrm{su}(d)$ is the
Lie algebra for the special unitary group $\mathrm{SU}(d)$.  This group
theoretical fact is highly relevant to the problem of characterizing
the geometry of quantum state
space~\cite{Mahler,Jakobczyk,Kim1,Byrd,Schirmer,Kim2,Dietz}.  However,
it will play no part in the considerations of this paper.  
}.  
 Let $\mathcal{B}$
be the convex subset consiting  of all $\hat{B} \in \mathrm{su}(d)$ for
which
$\hat{B}
\ge -1$.  Then a $d\times d$ matrix $\hat{\rho}$ is a density matrix if
and only if
\begin{equation}
\hat{\rho} =  \frac{1}{d} ( 1+\hat{B})
\end{equation}
for some $\hat{B} \in \mathcal{B}$.  We refer to $\mathcal{B}$ as the
Bloch body, and to its elements as Bloch vectors\footnote{What we are
calling Bloch vectors are of course matrices.  Some authors introduce a
standard basis for $\mathrm{su}(d)$ at this point and reserve the term
``Bloch vector'' for the components of $\hat{B}$ in that basis  (as has
been the long-standing practice in the $2$ dimensional case---see
Eq.~(\ref{eq:Bloch2D}) above).  However, it appears to us that this
makes the notation needlessly complicated. }.

It is convenient to define an inner product on $\mathrm{su}(d)$ by
\begin{equation}
\langle \hat{B}_1, \hat{B}_2 \rangle = \frac{1}{d(d-1)} \Tr (\hat{B}_1 
\hat{B}_2)
\label{eq:InnerPrdDef}
\end{equation}
for all $\hat{B }_1,  \hat{B}_2\in \mathrm{su}(d)$ (so $\langle
\hat{B}_1, \hat{B}_2 \rangle$ is just the Hilbert-Schmidt inner product
rescaled by the factor $\tfrac{1}{d(d-1)}$).  Let 
\begin{equation}
\|\hat{B}\| = \sqrt{\langle \hat{B}, \hat{B} \rangle}
\end{equation}
be the corresponding norm. 

If $d=2$ a vector $\hat{B}\in \mathrm{su}(d)$ is a Bloch vector if and
only if
$\|\hat{B}\|\le 1$.  Moreover, the corresponding density matrix is a pure
state if and only if $\|\hat{B}\|= 1$.  For
$d>2$ the situation is  more complicated. 
Let $\mathcal{B}_{\mathrm{I}}$ and $\mathcal{B}_{\mathrm{o}}$ be the balls
\begin{align}
\mathcal{B}_{\mathrm{i}}& = \Bigl\{\hat{B}\in\mathrm{su}(d)\colon
\|\hat{B}\|
\le
\frac{1}{d-1}\Bigr\}
\\
\mathcal{B}_{\mathrm{o}}& = \Bigl\{\hat{B}\in\mathrm{su}(d)\colon
\|\hat{B}\|
\le 1\Bigr\}
\end{align}
and let
\begin{align}
\mathcal{S}_{\mathrm{i}}& = \Bigl\{\hat{B}\in\mathrm{su}(d)\colon
\|\hat{B}\|
=
\frac{1}{d-1}\Bigr\}
\\
\mathcal{S}_{\mathrm{o}}& = \Bigl\{\hat{B}\in\mathrm{su}(d)\colon
\|\hat{B}\|
= 1\Bigr\}
\end{align}
be the bounding spheres.  Then~\cite{Harriman,Kim1,Kim2}
\begin{equation}
\mathcal{B}_{\mathrm{i}} \subseteq \mathcal{B}\subseteq
\mathcal{B}_{\mathrm{o}}
\label{eq:BigSmallBalls}
\end{equation}
It can further be shown~\cite{Harriman,Kim1,Kim2}  that
$\mathcal{B}_{\mathrm{i}}$ and 
$\mathcal{B}_{\mathrm{o}}$ are respectively the largest and smallest
balls centred on the origin for which this is true. Specifically:
\begin{enumerate}
\item If $r>1/(d-1)$ there exists $\hat{B} \in \mathrm{su}(d)$ such that
$\|\hat{B}\|=r$ and $\hat{B}\notin\mathcal{B}$.
\item If $0\le r \le 1$ there exists $\hat{B} \in \mathcal{B}$ such that
$\|\hat{B}\|=r$.
\end{enumerate} 
Moreover a Bloch vector $\hat{B}\in\mathcal{B}$ corresponds to a pure
state if and only if it has norm $=1$(\emph{i.e.}\ if and only if it $\in
\mathcal{B}\cap
\mathcal{S}_{\mathrm{o}}$).

It is worth noting that Bengtsson and
Ericsson~\cite{BengB} have proved a stronger result:  in any dimension
for which either a full set of MUBs (mutually unbiased bases) or an
SI(1)-POVM exist
$\mathcal{B}_{\mathrm{i}}$ is the largest
\emph{ellipsoid} which can be inscribed in $\mathcal{B}$.

If $d=2$ we have $\mathcal{B}_{\mathrm{i}} =\mathcal{B}
= \mathcal{B}_{\mathrm{o}}$ and
$\mathcal{B}\cap\mathcal{S}_{\mathrm{o}}=\mathcal{S}_{\mathrm{o}}$, so
the Bloch body has a very simple geometrical structure (it is just a
ball of radius $1$  centred on the origin, with the pure states
comprising the boundary).  For
$d>2$ these relations no longer hold, and the geometry  is much
harder to appreciate intuitively.  One gets some additional intuitive
feeling for  the geometry, at least in low dimension, by looking at the
$2$-dimensional sections of $\mathcal{B}$ which have been
calculated~\cite{Mahler,Jakobczyk,Kim1,Kim2} for
$d=3$ and
$4$. 

Let $\hat{B}$ be any vector 
$\in \mathcal{S}_{\mathrm{o}}$ (not necessarily a Bloch vector). An
immediate consequence of Eq.~(\ref{eq:BigSmallBalls}) is that $x \hat{B}
\in \mathcal{B}$ whenever
$|x| \le 1/(d-1)$. Kimura and Kossakowski~\cite{Kim2} have proved some
much stronger results. In the first place they have shown
\begin{theorem}
\label{thm:EvalThm1}
Let $\hat{B}$ be any vector 
$\in \mathcal{S}_{\mathrm{o}}$ (not necessarily a
Bloch vector).  Let $-m_{-}$  be the smallest
 eigenvalue of $\hat{B}$ and let $m_{+}$ be the largest (so
$-m_{-} \le \hat{B}
\le m_{+}$). Then
\begin{enumerate}
\item The quantities $m_{\pm}$ satisfy the inequalities 
\begin{align}
1 & \le m_{-} \le d-1
\\
\intertext{ and}
1 & \le m_{+} \le d-1
\end{align}
Moreover $m_{-}=1$  if and only if  $m_{+}=d-1$, and 
$m_{+}=1$  if and only if  $m_{-}=d-1$

\item $\hat{B}$  is a Bloch vector  (in fact the Bloch vector
corresponding to a pure state) if and only if
$m_{-}=1$.  Similarly $-\hat{B}$  is a Bloch vector  (in fact
the Bloch vector corresponding to a pure state) if and only if
$m_{+}=1$.
\end{enumerate}
\end{theorem}
\begin{proof}
See Kimura and
Kossakowski~\cite{Kim2}.
\end{proof}
Theorem~\ref{thm:EvalThm1} characterizes the vectors $\in
\mathcal{B}\cap\mathcal{S}_{\mathrm{o}}$ (\emph{i.e.}\ the Bloch vectors
corresponding to pure states) in terms of their eigenvalues.  The next
theorem relates the diameter of the Bloch body in the direction
$\hat{B}$ to the eigenvalues of $\hat{B}$.
\begin{theorem}
\label{thm:EvalThm2}
Let $\hat{B}$ and $m_{\pm}$ be as in the statement of
Theorem~\ref{thm:EvalThm1}, and let $x\in \mathbb{R}$.  Then $x\hat{B}\in
\mathcal{B}$  if and only if
\begin{equation}
-\frac{1}{m_{+}} \le x \le \frac{1}{m_{-}}
\end{equation}
\end{theorem}
\begin{proof}
See Kimura and
Kossakowski~\cite{Kim2}.
\end{proof}
\begin{remark}
As Kimura and Kossakowski point out, it follows
from Theorems~\ref{thm:EvalThm1} and~\ref{thm:EvalThm2}  that a point
where the boundary of
$\mathcal{B}$ touches the outer sphere
$\mathcal{S}_{\mathrm{o}}$ is always diametrically opposite to a point
where the boundary of $\mathcal{B}$ touches the inner sphere
$\mathcal{S}_{\mathrm{i}}$ (and conversely).
\end{remark}

We conclude this section by proving a theorem which shows that, instead
of considering the eigenvalues (as in Theorem~\ref{thm:EvalThm1}), one can
use the quantity
$\Tr(\hat{B}^3)$ to tell whether a vector $\hat{B}\in
\mathcal{S}_{\mathrm{o}}$ is the Bloch vector corresponding to a pure
state.  We first need to prove
\begin{lemma}
\label{lem:TrPCubedCond}
Let $\hat{P}$ be any $d\times d$ Hermitian matrix (not necessarily a
positive matrix).  Suppose
\begin{equation}
\Tr (\hat{P}^2) =1
\label{eq:TrPSquaredCond}
\end{equation}
Then
\begin{equation}
\Tr (\hat{P}^3) \le 1
\end{equation}
with equality if and only if $\hat{P}$ is a one dimensional projector.
\end{lemma}
\begin{remark}
It is \emph{not} assumed  that $\Tr(\hat{P})=1$.
\end{remark}
\begin{proof}
Let $\lambda_1,\lambda_2,\dots,\lambda_d$ be the eigenvalues of $\hat{P}$
(not necessarily distinct).  In view of Eq.~(\ref{eq:TrPSquaredCond})
\begin{equation}
\sum_{r=1}^{d} \lambda_r^2 = 1
\label{eq:TrPSquaredCondB}
\end{equation}
Define
\begin{equation}
\kappa = \sum_{r=1}^{d} |\lambda_r|^3
\end{equation}
It follows from Eq.~(\ref{eq:TrPSquaredCondB}) that $|\lambda_r|\le1$ for
all $r$, and consequently that $1-|\lambda_r|\ge 0$ for all $r$.  So
\begin{align}
1-\kappa & = \sum_{r=1}^{d} (\lambda_r^2-|\lambda_r|^3)
\nonumber
\\
& = \sum_{r=1}^{d} \lambda_r^2(1-|\lambda_r|)
\nonumber
\\
& \ge 0
\end{align}
with equality if and only if $\lambda_r^2 (1-|\lambda_r|) =0 $ for all
$r$.  Consequently
\begin{equation}
\kappa \le 1
\label{eq:kappaCondA}
\end{equation}
with equality if and only if
\begin{equation}
\lambda_r^2 (1-|\lambda_r|) =0 
\label{eq:kappaCondB}
\end{equation}
for all $r$.

It is now immediate that 
\begin{equation}
\Tr(\hat{P}^3) \le \bigl|\Tr(\hat{P}^3)\bigr| \le \kappa \le 1
\label{eq:TrPCubedCond}
\end{equation}

Suppose
\begin{equation}
\Tr(\hat{P}^3) =1
\end{equation}
Then it follows from Eq.~(\ref{eq:TrPCubedCond}) that $\kappa=1$ which
means, in view of Eqs.~(\ref{eq:kappaCondA}) and~(\ref{eq:kappaCondB}),
that $\lambda_r^2 (1-|\lambda_r|) =0$ for all $r$.  Consequently, for
each $r$, $|\lambda_r|=0$ or $1$.  The fact that $\sum_r \lambda_r^2=1$
then implies that $|\lambda_r|=1$ for exactly one value of $r$and $=0$
for all the others.  Since, by assumption, $\sum_r
\lambda_r^3=1$ we must actually have $\lambda_{r}=1$ for exactly one
value of $r$ and $=0$ for all the others---implying that $\hat{P}$ is a
one dimensional projector.

If, on the other hand, $\hat{P}$ is a one dimensional projector it is
immediate that $\Tr(\hat{P}^3)=1$.
\end{proof}
We are now in a position to prove our main result:
\begin{theorem}
\label{thm:TrCubedCondition}
Let $\hat{B}$ be any vector $\in \mathcal{S}_{\mathrm{o}}$ (not
necessarily a Bloch vector).  Then
\begin{enumerate}
\item The quantity  $\Tr(\hat{B}^3)$ satisfies the inequalities
\begin{equation}
 - d(d-1)(d-2) \le \Tr(\hat{B}^3) \le d (d-1)(d-2)
\label{eq:TrCubedCond}
\end{equation}
\item The upper bound in Inequalities~(\ref{eq:TrCubedCond}) is achieved
if and only if $\hat{B} \in \mathcal{B}\cap\mathcal{S}_{\mathrm{o}}$ (and
is therefore the Bloch vector corresponding to a pure state).
\item The lower bound in Inequalities~(\ref{eq:TrCubedCond}) is achieved
if and only if $-\hat{B} \in \mathcal{B}\cap\mathcal{S}_{\mathrm{o}}$ (and
is therefore the Bloch vector corresponding to a pure state).
\end{enumerate}
\end{theorem}
\begin{proof}
The fact that $\hat{B} \in \mathcal{S}_{\mathrm{o}}$ means
\begin{equation}
\Tr(\hat{B}^2) = d(d-1)
\label{eq:TrBSquaredCond}
\end{equation}
Define
\begin{equation}
\hat{P}_{\pm} = \frac{1}{d}(1\pm \hat{B})
\end{equation}
Then Eq.~(\ref{eq:TrBSquaredCond}) implies
\begin{equation}
\Tr(\hat{P}_{\pm}^2) = \frac{1}{d^2}\bigl(d+ \Tr(\hat{B}^2)\bigr)=1
\end{equation}
We may therefore use Lemma~\ref{lem:TrPCubedCond} to deduce
\begin{equation}
\frac{1}{d^3} \bigl(d+3 \Tr(\hat{B}^2)\pm\Tr(\hat{B}^3)\bigr)
=\Tr(\hat{P}_{\pm}^3) \le 1
\end{equation}
with equality if and only if $\hat{P}_{\pm}$ is a one dimensional
projector. In view of Eq.~(\ref{eq:TrBSquaredCond}) this means
\begin{equation}
\Tr(\hat{B}^3) \le d(d-1)(d-2)
\end{equation}
with equality if and only if $\hat{P}_{+}$ is a one dimensional projector,
and
\begin{equation}
\Tr(\hat{B}^3) \ge - d(d-1)(d-2)
\end{equation}
with equality if and only if $\hat{P}_{-}$ is a one dimensional projector.
But $\hat{P}_{+}$ is a one dimensional projector if and only if
$\hat{B}\in \mathcal{B}\cap\mathcal{S}_{\mathrm{0}}$, and $\hat{P}_{-}$ is
a one dimensional projector if and only if
$-\hat{B}\in \mathcal{B}\cap\mathcal{S}_{\mathrm{0}}$.  The claim is now
immediate.
\end{proof}
\section{Bloch Geometry and Tomography}
\label{sec:BlochAndToography}
The geometry of the Bloch body is intimately related to the problem of
devising measurement schemes which are, in some suitably defined sense,
tomographically optimal.  The connection works both ways.  On the one
hand knowledge of the geometry tells us what measurement schemes are
possible.  On the other hand a knowledge of possible measurement schemes
provides important insight into the geometry.  In this section we
summarize the Bloch geometrical aspects of two such measurement schemes: 
namely, schemes based on a full set of mutually unbiased bases or
MUBs~\cite{WoottersC,BengA,GrasslA,BengB,Zauner,Ivanovic,WoottersB,
WoottersA,Bandy,Pitt,KlappC,Archer,Durt,WoottersE,Planat}   and schemes
based on SI(1)-POVMs~\cite{Renes, FuchsA,WoottersC,BengA,GrasslA,BengB,
ApplebyB,GrasslB,KlappA,KlappB,Scott,Flammia,Kim} (or SIC-POVMs as they are often
called).  Much of the material in this section amounts to a review
of the relevant parts of Bengtsson~\cite{BengA} and Bengtsson and
Ericsson~\cite{BengB}, but using a slightly different terminology and
notation.

We begin with the case of  a full set of MUBs.  Suppose one has a large
number of copies  of a $d$-dimensional quantum system, all presumed to be
in the same quantum state.  Suppose one takes a fixed von Neumann
measurement having $d$ distinct outcomes, and performs it on many copies
of the system.  Suppose one then identifies the relative frequencies
obtained with the corresponding probabilities.  This will give one
$d$ probabilities
$p_1,p_2,\dots,p_d$.  Taking into account the normalisation condition
$\sum_{r=1}^d p_r =1$ this means one has $d-1$ independent numbers.  On
the other hand a full specification of the quantum state requires $d^2-1$
independent numbers.  It follows that if one wants to perform
tomography using only von Neumann measurements one needs to divide the set
of copies of the system into a  minimum of
$d+1$ subsets, and to perform different von Neumann measurements on the
copies belonging to different subsets.  We will refer to a measurement
scheme based on the minimum number of $d+1$ different von Neumann
measurements, each having $d$ distinct outcomes, as a \emph{minimal Von
Neumann scheme}.

The question now arises:  what is the best way of choosing the $d+1$
different measurements in a minimal von Neumann scheme?  Let
$\hat{P}^{r}_1,\hat{P}^{r}_2,\dots,\hat{P}^{r}_d$ be the $d$
orthogonal, one dimensional projectors describing the $r^{\mathrm{th}}$
measurement and let $\hat{B}^{r}_1,\hat{B}^{r}_2,\dots,\hat{B}^{r}_d$ be
the corresponding Bloch vectors.  So
\begin{equation}
\hat{P}^r_a = \frac{1}{d}(1+\hat{B}^r_a)
\end{equation}
for all $a$, $r$.  Notice that, whereas in Section~\ref{sec:BlochBody} we
used Bloch vectors to describe quantum states, now we are using them to
describe quantum measurements.  Notice also that the fact that the
$\hat{P}^{r}_a$ are all one dimensional projectors means that  the
vectors $\hat{B}^{r}_a$ all lie on
$\mathcal{B}\cap\mathcal{S}_{\mathrm{o}}$. 

The orthonormality condition
\begin{equation}
\Tr(\hat{P}^r_a\hat{P}^r_b) = \delta_{ab}
\end{equation} 
together with Eq.~(\ref{eq:InnerPrdDef}) implies
\begin{equation}
\langle \hat{B}^r_a,\hat{B}^r_b\rangle = 
\begin{cases}
1 \quad & a = b \\
-\frac{1}{d-1}\quad & a\neq b
\end{cases}
\end{equation}
from which one sees that for each $r$ the Bloch vectors
$\hat{B}^r_1,\hat{B}^r_2,\dots, \hat{B}^r_d$ are the vertices of a
regular $d-1$ dimensional simplex.  
So the $d+1$ families of orthogonal projectors define $d+1$ regular
simplices, each having its vertices in
$\mathcal{B}\cap\mathcal{S}_{\mathrm{o}}$.  One might guess, and detailed
calculation confirms~\cite{WoottersA,ApplebyA}, that the optimal choice
from a tomographic point of view is to choose the projectors in such a
way that the polytope formed by all $d^2+d$ Bloch vectors has maximal
volume.  This is achieved if the $d+1$ different simplices are mutually
orthogonal:
\begin{equation}
\langle \hat{B}^r_a,\hat{B}^s_b\rangle = 0
\end{equation}
for all $a$, $b$ and $r\neq s$.

This condition is often stated in a slightly different form.  Suppose we
choose vectors
$|\psi^r_a\rangle \in
\mathcal{H}$  such that
$\hat{P}^r_a =|\psi^r_a\rangle\langle\psi^r_a|$ (so for each $r$ the
set
$|\psi^r_1\rangle, |\psi^r_2\rangle,\dots, | \psi^r_d\rangle $ is an
orthonormal basis for $\mathcal{H}$).  Then the requirement that the
simplices corresponding to different bases be mutually orthogonal is
equivalent to the requirement that
\begin{equation}
\bigl|\langle \psi^r_a | \psi^s_b \rangle\bigr| = \frac{1}{\sqrt{d}}
\end{equation}
for all $a$, $b$ and  $r\neq s$.  A family of orthonormal bases for which
this condition is satisfied is said to be \emph{mutually unbiased}.

The question now arises:  do families of $d+1$ mutually unbiased bases
(MUBs) actually exist?  This is a difficult geometrical problem.  As
Bengtsson and Ericsson~\cite{BengA,BengB} have noted, what makes it hard
is, in essence, the fact that
$\mathcal{B}\cap\mathcal{S}_{\mathrm{o}}$ has a much lower dimension than
$\mathcal{S}_{\mathrm{o}}$.  Consider, for instance, the case
$d=3$.  In that case the problem is to orientate a set of $4$ mutually
orthogonal equilateral triangles in such a way that all $12$ vertices lie
in
$\mathcal{B}\cap \mathcal{S}_{\mathrm{o}}$.  It is very easy, almost
trivial,  to construct  a family of $4$ mutually orthogonal equilateral
triangles with vertices on the $7$ dimensional sphere
$\mathcal{S}_{\mathrm{o}}$.  The difficult part is then to rotate them so
that every vertex lies on the $4$ dimensional subspace $\mathcal{B}\cap
\mathcal{S}_{\mathrm{o}}$.  As it happens the problem has been solved for
$d=3$, and also for every other  dimension which is the power of a prime
number~\cite{Ivanovic,WoottersA}.  But for values of
$d$ which are not prime powers the question is still 
open~\cite{WoottersC,BengA,GrasslA,BengB,Zauner,Ivanovic,WoottersB,
WoottersA,Bandy,Pitt,KlappC,Archer,Durt,WoottersE,Planat}.
So we have here  an important physical problem the solution to
which depends on gaining a better understanding of the geometry of the
Bloch body.  

Let us now turn to a different measurement scheme.  Suppose that, instead
of using $d+1$ different von Neumann measurements, we wanted to use a
single POVM measurement.  The POVM would  obviously need to have the
property that specifying the probability of each of the distinct
outcomes fixes the quantum state.  Such a POVM is said to be
\emph{informationally
complete}~\cite{Prugo,Schroeck,BuschA,BuschB,Caves,FuchsB,dAriano}.   As
we remarked earlier, a complete specification of the quantum state
requires the specification of
$d^2-1$ independent numbers.  Taking into account the normalisation
condition (the fact that the probabilities must sum to unity) this means
that an informationally complete POVM  must have at least $d^2$
distinct outcomes.  We will say that a POVM is minimal informationally 
complete
 if it has
precisely this minimum number of
$d^2$ distinct outcomes.  The question we have then to answer is:  which
minimal informationally complete POVMs are tomographically optimal?  As
with the MUB problem, the answer to this question depends on achieving a
better understanding of the geometry of the Bloch body.

Let $\hat{E}_1,\hat{E}_2,\dots,\hat{E}_{d^2}$ be an arbitrary POVM having
$d^2$ distinct elements.  Define 
\begin{equation}
t_r =\Tr(\hat{E}_r)
\end{equation}
We may assume that $\hat{E}_r\neq 0$ and consequently $t_r\neq 0$ for all
$r$ (otherwise the POVM would effectively reduce to one having fewer
than $d^2$ elements).  It follows that for each $r$ the operator
$(1/t_r)\hat{E}_r$ is a density matrix.  We may therefore write, for all
$r$,
\begin{equation}
\hat{E}_r = \frac{t_r}{d} (1+ \hat{B}_r)
\label{eq:POVMtermsB}
\end{equation}
where $\hat{B}_r\in \mathcal{B}$.  The fact that
$\sum_{r=1}^{d^2}\hat{E}_r = 1$ implies
\begin{align}
\sum_{r=1}^{d^2} t_r & = d
\label{eq:trSum}
\\
\intertext{and}
\sum_{r=1}^{d^2} t_r \hat{B}_r & = 0
\label{eq:trBSum}
\end{align}
It is easily seen that the POVM is informationally complete if and only
if the Bloch vectors $\hat{B}_r$ span $\mathrm{su}(d)$.  This in turn
will be true if and only if the vectors $\hat{B}_r$ are the vertices of a
$d^2-1$ dimensional simplex (typically an irregular simplex) having
non-zero volume.  One might guess, and detailed calculation
confirms~\cite{ApplebyA}, that the POVM would be optimal from a
tomographic point of view if we could arrange that (a)
 the simplex is regular and (b) the vertices all lie on
$\mathcal{B}\cap\mathcal{S}_{\mathrm{o}}$ (because the volume
of the simplex would then be maximal).  In other words we would like to
arrange that
\begin{equation}
\langle\hat{B}_r,\hat{B}_s\rangle
=
\begin{cases} 1 \quad & r =s \\
-\frac{1}{d^2-1} \quad & r \neq s
\end{cases}
\label{eq:SI1BlochCond}
\end{equation}
In that case
\begin{equation}
\sum_{r=1}^{d^2} \langle  \hat{B}_r, \hat{B}_s \rangle = 0
\end{equation}
for all $s$. Since the vectors $\hat{B}_s$ span $\mathrm{su}(d)$ this
means 
\begin{equation}
\sum_{r=1}^{d^2} \hat{B}_r=0
\label{eq:BSum}
\end{equation}   
Eqs.~(\ref{eq:trSum}), (\ref{eq:trBSum}) and~(\ref{eq:BSum}), taken in
conjunction with the fact that the $d^2$ vectors $\hat{B}_r$ span the
$d^2-1$ dimensional space $\mathrm{su}(d)$, then imply
\begin{equation}
t_1=t_2=\dots = t_{d^2} = \frac{1}{d}
\end{equation}
so that the POVM elements take the form
\begin{equation}
\hat{E}_r = \frac{1}{d^2} (1+\hat{B}_r)
\end{equation}
Since the $\hat{B}_r$ all belong to
$\mathcal{B}\cap\mathcal{S}_{\mathrm{o}}$ we may alternatively write
\begin{equation}
\hat{E}_r = \frac{1}{d} \hat{P}_r
\label{eq:ETermsP}
\end{equation}
where the $\hat{P}_r$ are a family of one dimensional projectors
satisfying
\begin{equation}
\Tr(\hat{P}_r\hat{P}_s)
=
\begin{cases} 1 \quad & r = s \\ \frac{1}{d+1} \quad & r\neq s
\end{cases}
\label{eq:SI1ProjCond}
\end{equation}
The converse is also 
true\footnote{To 
   see this note that Eq.~(\ref{eq:SI1ProjCond})
   implies that the corresponding Bloch vectors satisfy
   Eq.~(\ref{eq:SI1BlochCond}).  It follows that the Bloch vectors span
   $\mathrm{su}(d)$ (because if $M$ is the $(d^2-1)\times (d^2-1)$ matrix
   with elements $M_{r s} = \langle \hat{B}_r, \hat{B}_s\rangle$ for
   $r,s=1,2,\dots, d^2-1$ then
   $\Det M = d^{2(d^2-2)}/(d^2-1)^{d^2-1}\neq0$ ) and consequently that
   $\sum_{r=1}^{d^2}\hat{B}_r = 0$ (by the same argument that led to
   Eq.~(\ref{eq:BSum})).  The claim is now immediate.
}:   
if
$\hat{P}_1,
\hat{P}_2,\dots ,
\hat{P}_{d^2}$ is any family of one dimensional projectors satisfying
Eq.~(\ref{eq:SI1ProjCond}) then $\frac{1}{d}\hat{P}_1,
\frac{1}{d}\hat{P}_2,\dots ,
\frac{1}{d}\hat{P}_{d^2}$ is an informationally 
complete POVM.

A POVM  which satisfies the defining Eq.~(\ref{eq:SI1BlochCond})
(equivalently: a POVM which  is rank 1 and which satisfies 
Eqs.~(\ref{eq:ETermsP}) and~(\ref{eq:SI1ProjCond})) is usually referred
to  as a SIC-POVM (symmetric informaationally complete POVM).  It
appears to us that this terminology is unsatisfactory as, besides being
symmetric and informationally complete, POVMs of the type in question
are also rank 1.  As we will see in Section~\ref{sec:SymmGeneral}, there
do exist POVMs which are symmetric and informationally complete but not
rank 1.  We therefore suggest that POVMs of the type in question would
be better described as SI(1)-POVMs (``S'' for symmetric, ``I'' for
informationally complete, ``1'' for rank 1). The larger class of POVMs,
which are symmetric and informationally complete but not necessarily
rank 1, we will refer to as SI-POVMs.

Do SI(1)-POVMs exist?  This is a difficult geometrical problem.  Moreover,
it is difficult for essentially the same reason that the MUB problem is
difficult~\cite{BengA,BengB}:  namely, the submanifold
$\mathcal{B}\cap\mathcal{S}_{\mathrm{o}}$ has much lower dimension than
the sphere $\mathcal{S}_{\mathrm{o}}$ if
$d>2$.  It is easy to consruct a regular $d^2-1$ dimensional simplex with
vertices in the $d^2-2$ dimensional sphere $\mathcal{S}_{\mathrm{o}}$, 
but very hard then to rotate the simplex so that every vertex lies
in the $2 (d-1)$ dimensional subspace
$\mathcal{B}\cap\mathcal{S}_{\mathrm{o}}$ (except, of course, when
$d=2$).  Moreover the difficulty increases with increasing $d$ (because
$2(d-1)/(d^2-2)\to 0$ as $d\to \infty$).  
SI(1)-POVMs have been constructed
analytically~\cite{Renes,GrasslA,ApplebyB,GrasslB,Hoggar,Zauner} in
dimensions 2 to 10 inclusive, and  in dimensions 12, 13 and 19.  They
have been constructed numerically~\cite{Renes} in
 dimensions 5 to 45 inclusive.  It is  an open question whether they
exist in dimensions $>45$.

Most (not all) of the SI(1)-POVMs which have been constructed to date are
covariant under the action of the 
Weyl-Heisenberg group (or generalized Pauli group, as it is sometimes
called).  For a summary of the pertinent facts concerning this group see
Appendix~\ref{sec:WHGroup}.

Let $\mathbb{Z}_d^2$ be the set of integer pairs $\mathbf{p} =(p_1,p_2)$
such that $0\le p_1,p_2 \le d-1$, and for each $\mathbf{p} \in
\mathbb{Z}_d^2$ let $\hat{D}_{\mathbf{p}}$ be the corresponding
Weyl-Heisenberg displacement operator, as defined by
Eq.~(\ref{eq:DOpDefinition}).  Let $\hat{B}$ be any Bloch vector $\in
\mathcal{B}\cap\mathcal{S}_{\mathrm{o}}$.  Then the fact that the
$\hat{D}_{\mathbf{p}}$ are unitary means that for each $\mathbf{p}$
\begin{equation}
\hat{B}_{\mathbf{p}} =\hat{D}^{\vphantom{\dagger}}_{\mathbf{p}}
\hat{B}\hat{D}^{\dagger}_{\mathbf{p}}
\end{equation}
also belongs to $
\mathcal{B}\cap\mathcal{S}_{\mathrm{o}}$.  Suppose that the
$\hat{B}_{\mathbf{p}}$ constitute a regular simplex:
\begin{equation}
\langle\hat{B}_{\mathbf{p}},\hat{B}_{\mathbf{q}}\rangle
=
\begin{cases} 1 \quad & \mathbf{p} =\mathbf{q} \\
-\frac{1}{d^2-1} \quad & \mathbf{p} \neq \mathbf{q}
\end{cases}
\label{eq:WHSimplexCond}
\end{equation}
Then the corresponding SI(1)-POVM is said to be Weyl-Heisenberg covariant.
\section{SI-POVMs in General}
\label{sec:SymmGeneral}
In the last section we discussed SI(1)-POVMs:  POVMs which are not only
symmetric and informationally complete but also rank-1 (so that each
element of the POVM is proportional to a one dimensional projector).  We
now want to broaden the discussion, and consider POVMs which, though
symmetric and informationally complete, are not necessarily rank 1.

Consider an arbitrary POVM $\hat{E}_1, \hat{E}_2, \dots \hat{E}_n$ defined
on a $d$ dimensional Hilbert space. Without loss of generality it may be
assumed that 
$\hat{E}_r
\neq 0$ for all
$r$.  We saw in the last section that we can write 
\begin{equation}
\hat{E}_r = \frac{t_r}{d} (1+ \hat{B}_r)
\label{eq:POVMtermsB2}
\end{equation}
where $\hat{B}_r \in \mathcal{B}$ for all $r$,  where $t_r > 0$ for all
$r$, and where
\begin{align}
\sum_{r} t_r & = d
\label{eq:trSum2}
\\
 \sum_{r} t_r \hat{B}_r & = 0
\label{eq:trBSumB}
\end{align}
Conversely, if we have a set of Bloch vectors $\hat{B}_r$ and
positive numbers $t_r$ satisfying these conditions then
Eq.~(\ref{eq:POVMtermsB2}) defines a POVM.

We say that the POVM is informationally complete if   the probabilities
$\Tr(\hat{\rho} \hat{E}_r)$ completely specify an arbitrary density matrix
$\hat{\rho}$.   
We say that it is symmetric if 
\begin{equation}
\Tr (\hat{E}_r \hat{E}_s) = \alpha + \beta \delta_{r s}
\label{eq:SymmCond}
\end{equation}
for all $r, s$ and fixed numbers $\alpha, \beta$.

We then have the following theorem:
\begin{theorem}
\label{thm:SIPOVMChar}
Let $\hat{E}_1, \hat{E}_2, \dots , \hat{E}_n$ be a POVM having $n$
elements (all non-zero) defined on a $d$-dimensional Hilbert space.  The
POVM is symmetric and informationally complete if and only if
\begin{enumerate}
\item $n=d^2$.
\item The POVM elements are of the form
\begin{equation}
\hat{E}_r = \frac{1}{d^2} (1+\hat{B}_r)
\end{equation}
where the Bloch vectors $\hat{B}_r$ satisfy
\begin{equation}
\langle \hat{B}_r , \hat{B}_s\rangle
=
\begin{cases} \kappa^2 \quad & r=s \\
-\frac{\kappa^2}{d^2-1} \quad & r\neq s
\end{cases}
\label{eq:BrBsProduct}
\end{equation}
with $0 < \kappa \le 1$.
\end{enumerate}
\end{theorem}
\begin{remark} We will refer to $\kappa$ as the efficiency parameter as it
determines the volume  of the regular simplex spanned by the Bloch vectors
$\hat{B}_r$, and consequently the efficiency of the POVM for tomographic
purposes~\cite{ApplebyA}.  The POVM is maximally efficient if and only if
$\kappa =1$ in which case it is rank one (an SI(1)-POVM in the terminology
explained in the last section).
\end{remark}
\begin{proof}
We first prove necessity.  Suppose the POVM is symmetric and
informationally complete.  We can write it in the form  
\begin{equation}
\hat{E}_r = \frac{t_r}{d} (1+ \hat{B}_r)
\label{eq:POVMtermsB3}
\end{equation}
for  Bloch vectors $\hat{B}_r$ and positive numbers $t_r$ satisfying
Eqs.~(\ref{eq:trSum2}) and~(\ref{eq:trBSumB}).  The symmetry condition
Eq.~(\ref{eq:SymmCond}) then implies
\begin{equation}
t_r = \Tr(\hat{E}_r) = \sum_{s=1}^{n} \Tr(\hat{E}_r\hat{E}_s) 
= n \alpha + \beta
\end{equation}
for all $r$.  In view of  Eq.~(\ref{eq:trSum2})
this means
\begin{align}
t_r & = \frac{d}{n} 
\\
\intertext{for all $r$, and consequently}
\alpha & = \frac{d - n \beta }{n^2}  
\end{align}
Using these results, Eq.~(\ref{eq:POVMtermsB3}) and the symmetry condition
Eq.~(\ref{eq:SymmCond}) we deduce
\begin{equation}
\langle \hat{B}_r, \hat{B}_s\rangle = -\frac{n \beta}{d (d-1)}+\frac{n^2
\beta}{d(d-1)} \delta_{r s}
\label{eq:BrBsProductB}
\end{equation}
The fact that the $\hat{B}_r$ are  Bloch vectors means $\langle
\hat{B}_r, \hat{B}_r\rangle \le 1$.  We must also have $\langle
\hat{B}_r, \hat{B}_r\rangle > 0$ (because otherwise $\hat{E}_r =
\frac{1}{n}$ for all $r$, in which case the POVM would not be
informationally complete).  Consequently
\begin{equation}
0 < \beta \le  \frac{d(d-1)}{n(n-1)}
\label{eq:BetaLimits}
\end{equation}
Let $\hat{M}$ be the $n\times n$ matrix with elements $\hat{M}_{r
s}=\langle \hat{B}_r , \hat{B}_s \rangle$.  Since the POVM is
informationally complete the Bloch vectors $\hat{B}_r$ must span the
$d^2-1$ dimensional space $\mathrm{su}(d)$.  So $\hat{M}$ must have rank
$d^2-1$.  On the other hand
\begin{equation}
\Det (\hat{M}-\lambda) = -\lambda \Bigl(\frac{n^2
\beta}{d(d-1)}-\lambda\Bigr)^{n-1}
\end{equation}
It follows from this that $\hat{M}$ has  $n-1$ non-zero eigenvalues
(since we have shown that
$\beta >0$).   However, the fact that $\hat{M}$ is rank $d^2-1$ means
that it must have  $d^2-1$ non-zero eigenvalues.  We conclude that
$n=d^2$.  Making the substitutions $n=d^2$ and 
$\beta=\frac{\kappa^2}{d(d+1)}$ in Eq.~(\ref{eq:BrBsProductB}) we obtain
Eq.~(\ref{eq:BrBsProduct}). Moreover, it follows
from Eq.~(\ref{eq:BetaLimits}) that $0<\kappa \le 1$.

Having proved necessity, it remains to prove sufficiency.  Suppose
$\hat{B}_1, \hat{B}_2, \dots , \hat{B}_{d^2}$ are Bloch vectors
satisfying  Eq.~(\ref{eq:BrBsProduct}).  Let $\hat{M}$ be the $d^2\times
d^2$ matrix with elements $\hat{M}_{r s} = \langle \hat{B}_r,
\hat{B}_s\rangle$.  Then
\begin{equation}
\Det(\hat{M}-\lambda)= - \lambda \Bigl(\frac{\kappa^2 d^2}{d^2-1}- \lambda
\Bigr)^{d^2-1}
\end{equation}
Since, by assumption, $\kappa > 0$ it follows that $\hat{M}$ has $d^2-1$
non-zero eigenvalues, and is therefore rank $d^2-1$.   Consequently the
Bloch vectors span the $d^2-1$ dimensional vector space $\mathrm{su}(d)$. 

Eq.~(\ref{eq:BrBsProduct}) also implies
\begin{equation}
\Bigl< \Bigl(\sum_{s=1}^{d^2} \hat{B}_s \Bigr) , \hat{B}_r\Bigr> = 0
\end{equation}
for all $r$.  Since the $\hat{B}_r$ span $\mathrm{su}(d)$ we deduce
\begin{equation}
\sum_{s=1}^{d^2} \hat{B}_s =0
\end{equation}
It follows from this that if we define
\begin{equation}
\hat{E}_r = \frac{1}{d^2} \bigl(1+ \hat{B}_r  \bigr)
\end{equation}
the operators $\hat{E}_1,\hat{E}_2, \dots, \hat{E}_{d^2}$ constitute a
POVM.  The fact that the $\hat{B}_r$ span $\mathrm{su}(d)$ means the POVM
is informationally complete.  The fact that the POVM is symmetric is
immediate.
\end{proof}

We noted in the last section that the existence problem for SI(1)-POVMs is
hard, and still unsolved for dimensions $>45$.   But if one relaxes the
demand that the POVM be rank 1, and simply looks for an SI-POVM of
arbitrary rank, the problem becomes much easier.  

To construct an SI-POVM of arbitrary rank all we have to do is construct  a
regular simplex in $\mathrm{su} (d)$  with its vertices all on
$\mathcal{S}_{\mathrm{o}}$ (since $\mathcal{S}_{\mathrm{o}}$ is a sphere
such simplices are guaranteed to exist).  Let $\hat{B}_1, \hat{B}_2,\dots
\hat{B}_{d^2}$ be the vertices.  Then
\begin{equation}
\langle\hat{B}_r , \hat{B}_s\rangle 
= \begin{cases} 1 \quad & \text{if $r=s$}
\\
-\frac{1}{d^2-1} \quad & \text{otherwise}
\end{cases}
\end{equation}
If the $\hat{B}_r$ were Bloch vectors this would give us an SI(1)-POVM. 
However, if the simplex is chosen at random they are very unlikely to  be
Bloch vectors (because the manifold
$\mathcal{B}\cap
\mathcal{S}_{\mathrm{o}}$ has much lower dimension than
$\mathcal{S}_{\mathrm{o}}$).  Nevertheless, we can still use them to
construct an SI-POVM by shrinking the simplex  until the vertices are all
in $\mathcal{B}$.  In fact, let
$- m_{r -}$ be the smallest eigenvalue of $\hat{B}_r$.  It follows from
Theorem~\ref{thm:EvalThm1} that $1 \le m_{r -} \le d-1$ for all $r$. Now
define
\begin{equation}
\kappa = \min_{1\le r \le d^2} \Bigl(\frac{1}{m_{r-}}
\Bigr)
\end{equation}
We have $\frac{1}{d-1}\le \kappa \le 1$. Moreover, it follows from
Theorem~\ref{thm:EvalThm2} that $\hat{B}'_r=\kappa \hat{B}_r \in
\mathcal{B}$ for all $r$.  By construction
\begin{equation}
\langle\hat{B}'_r , \hat{B}'_s\rangle 
= \begin{cases} \kappa^2 \quad & \text{if $r=s$}
\\
-\frac{\kappa^2}{d^2-1} \quad & \text{otherwise}
\end{cases}
\end{equation}
So we can use Theorem~\ref{thm:SIPOVMChar} to deduce that the POVM with
elements
\begin{equation}
\hat{E}_r = \frac{1}{d^2} (1+ \hat{B}'_r)
\end{equation}
is symmetric, informationally complete with efficiency parameter $=\kappa$.

The argument just given shows that in every dimension $d$ there exists an
SI-POVM with efficiency parameter $\ge \frac{1}{d-1}$.  We will see in
Section~\ref{sec:WigPOVM} that at least when $d$ is odd it is possible to
considerably improve on that.

\section{SI-POVMs which are Weyl-Heisenberg Covariant}
\label{sec:WHCovariantSIPOVMs}
In Section~\ref{sec:SI1Construct} we will discuss the bearing of the above
results on the really difficult problem: 
\emph{i.e.}\ the problem of constructing POVMs which are, not merely
symmetric and informationally complete, but also rank 1 (have efficiency
parameter $=1$).  In preparation for that we first need to  prove a result
concerning SI-POVMs (with efficiency parameter not necessarily $=1$) which
are covariant under the Weyl-Heisenberg group.

We begin with a definition.  Let $\hat{B} \in
\mathcal{S}_{\mathrm{o}}$  (we do not assume that $\hat{B}$ is a
Bloch vector), and for  each $\mathbf{p} \in \mathbb{Z}^2_d$ let 
$\hat{B}_{\mathrm{p}}=\hat{D}^{\vphantom{\dagger}}_{\mathbf{p}}\hat{B}
\hat{D}^{\dagger}_{\mathbf{p}}$ (where
$\mathbb{Z}_d^2$ and
$\hat{D}_{\mathbf{p}}$ are as defined in Appendix~\ref{sec:WHGroup}).  We
say that $\hat{B}$ is the
generating vector for a Weyl-Heisenberg covariant regular simplex if
\begin{equation}
\langle \hat{B},\hat{B}_{\mathbf{p}} \rangle =
\begin{cases} 1 \quad & \text{if $\mathbf{p}=(0,0)$} \\
-\frac{1}{d^2-1} \quad & \text{otherwise}
\end{cases}
\end{equation}
It is easily seen that if that is the case
\begin{equation}
\langle \hat{B}_{\mathbf{p}},\hat{B}_{\mathbf{q}} \rangle =
\begin{cases} 1 \quad & \text{if $\mathbf{p}=\mathbf{q}$} \\
-\frac{1}{d^2-1} \quad & \text{otherwise}
\end{cases}
\end{equation}
meaning that the vectors $\mathbf{B}_{\mathbf{p}}$ are the vertices of a
regular simplex.

We now have the following lemma:
\begin{lemma}
A vector $\hat{B} \in
\mathcal{S}_{\mathrm{o}}$ is the generating vector for a Weyl-Heisenberg
covariant regular simplex if and only if
\begin{equation}
\hat{B} = \frac{1}{\sqrt{d+1}}
\sum_{\mathbf{q} \in (\mathbb{Z}_d^2)^{*}} e^{i
\theta_{\mathbf{q}}}
\hat{D}_{\mathbf{q}}
\label{eq:WHGeneratorTermsPhases}
\end{equation}
for any set of real numbers $\theta_{\mathbf{q}}$ satisfying the condition
$  e^{i
\theta_{\bar{\mathbf{q}} }}= s_{-\mathbf{q}} e^{-i
\theta_{\mathbf{q}}} $ (where $(\mathbb{Z}_d^2)^{*}$,  $s_{-\mathbf{q}}$
and
$\bar{\mathbf{q}}$ are as defined in Appendix~\ref{sec:WHGroup}).
\end{lemma}
\begin{proof}
We know from Eq.~(\ref{eq:sudExpansion}) that any  vector $\hat{B} \in
\mathcal{S}_{\mathrm{o}}$  can be written
\begin{equation}
\hat{B} = 
\sum_{\mathbf{q} \in (\mathbb{Z}_d^2)^{*}} c_{\mathbf{q}}
\hat{D}_{\mathbf{q}}
\end{equation}
where the expansion coefficients $c_{\mathbf{q}}= (1/d)
\Tr(\hat{D}^{\dagger}_{\mathbf{q}} \hat{B})$ satisfy the condition
$ c_{\bar{\mathbf{q}} } = s^{\vphantom{*}}_{-\mathbf{q}}
c^{*}_{\mathbf{q}} 
$.  By a straightforward application of Eq.~(\ref{eq:DCompositionRule}) we
find
\begin{equation}
\hat{B}_{\mathbf{p}} = 
\sum_{\mathbf{q} \in (\mathbb{Z}_d^2)^{*}}\tau^{2 \langle \mathbf{p},
\mathbf{q} \rangle}  c_{\mathbf{q}} 
\hat{D}_{\mathbf{q}}
\end{equation}
In view of Lemma~\ref{thm:TraceFormulae} in the Appendix it follows
\begin{align}
\langle \hat{B},\hat{B}_{\mathbf{p}} \rangle
& = \frac{1}{(d-1)}
\sum_{\mathbf{q} \in (\mathbb{Z}_d^2)^{*}} |c_{\mathbf{q}}|^2
\tau^{2\langle \mathbf{p}, \mathbf{q} \rangle}
\label{eq:BBpScalarProduct}
\end{align}
Suppose now that $|c_{\mathbf{q}}|=1/\sqrt{d+1}$ for all non-zero
$\mathbf{q}$.  Then Eq.~(\ref{eq:BBpScalarProduct}) implies
\begin{equation}
\langle \hat{B},\hat{B}_{\mathbf{p}} \rangle
 =
\frac{1}{d^2-1}\Bigl(-1+ \sum_{\mathbf{q}  \in \mathbb{Z}_d^2}
\tau^{2\langle \mathbf{p}, \mathbf{q}\rangle}
\Bigr)
 = \frac{1}{d^2-1} \bigl( -1 + d^2 \delta_{ \mathbf{p}\boldsymbol{
0}}\bigr)
\label{eq:WHCovSufficiency}
\end{equation}
So $\hat{B}$ is the generating vector for a Weyl-Heisenberg covariant
regular simplex.

To prove necessity, suppose that  Eq.~(\ref{eq:WHCovSufficiency}) is
satisfied.  Using the fact that
\begin{equation}
\sum_{\mathbf{p}\in \mathbb{Z}_d^2} \tau^{2\langle \mathbf{p} ,\mathbf{q}-
\mathbf{r}
 \rangle} = d^2 \delta_{\mathbf{q}\mathbf{r}}
\end{equation}
for all $\mathbf{q},\mathbf{r}\in\mathbb{Z}_d^2$ to invert
Eq.~(\ref{eq:BBpScalarProduct})  one finds
$|c_{\mathbf{q}}|=1/\sqrt{d+1}$ for all non-zero $\mathbf{q}$.
\end{proof}
This lemma gives us an easy way to construct SI-POVMs.  Simply choose an
arbitrary set of phases  $e^{i\theta_{\mathbf{q}}}$ satisfying the
condition
$e^{i
\theta_{\bar{\mathbf{q}} }} = s_{-\mathbf{q}} e^{-i
\theta_{\mathbf{q}}} $ and construct the vector $\hat{B}$ specified
by Eq.~(\ref{eq:WHGeneratorTermsPhases}).  Let $-1/\kappa$ be the
minimum eigenvalue of $\hat{B}$.  It follows from
Theorem~\ref{thm:EvalThm1} that $1/(d-1) \le \kappa \le 1$.  Moreover
$-1/\kappa$ is also the minimum eigenvalue of $\hat{B}_{\mathbf{p}}$ for
all
$\mathbf{p}$.  So it follows from 
Theorem~\ref{thm:EvalThm2}  that the operators 
\begin{equation}
\hat{E}_{\mathbf{p}} = \frac{1}{d^2} (1 + \kappa \hat{B}_{\mathbf{p}})
\label{eq:WHCovSIPOVM}
\end{equation}
consitute a POVM.  By construction the POVM is SI, Weyl-Heisenberg
covariant, and has efficiency parameter $\kappa\ge 1/(d-1)$.
\section{Construction of SI(1)-POVMs}
\label{sec:SI1Construct}
Of course, what we would really like to do is to construct a POVM which is,
not merely symmetric and informationally complete, but also rank 1.  The
POVM defined by Eq.~(\ref{eq:WHCovSIPOVM}) will be rank 1, with efficiency
parameter
$\kappa=1$, if and only if $\hat{B}$ is a Bloch vector.  The question
therefore arises:  how do we choose the phases in 
Eq.~(\ref{eq:WHGeneratorTermsPhases}) so as to ensure that that is the
case?

We can answer that question by appealing to
Theorem~\ref{thm:TrCubedCondition}.  The vector $\hat{B}$ in
Eq.~(\ref{eq:WHGeneratorTermsPhases}) is on the sphere
$\mathcal{S}_{\mathrm{o}}$.  So Theorem~\ref{thm:TrCubedCondition} tells us
that
\begin{equation}
\Tr(\hat{B}^3) \le d (d-1)(d-2)
\end{equation}
with equality if and only if $\hat{B}$ is a Bloch vector. In terms of the
phases on the right hand side of Eq.~(\ref{eq:WHGeneratorTermsPhases}) the
condition reads (using Lemma~\ref{thm:TraceFormulae} in the Appendix)
\begin{equation}
\sum_{\mathbf{p},\mathbf{q},\mathbf{p}\oplus\mathbf{q} \in
(\mathbb{Z}_d^2)^{*}} s_{\mathbf{p}+\mathbf{q}}
\tau^{\langle\mathbf{p},\mathbf{q}
\rangle}
e^{i(\theta_{\mathbf{p}}+\theta_{\mathbf{q}}-
\theta_{\mathbf{p}\oplus\mathbf{q}})}
\le (d-1)(d-2)(d+1)^{\frac{3}{2}}
\label{eq:PhaseExtremCond}
\end{equation}
with equality if and only if $\hat{B}$ is a Bloch vector.

This gives us an extremal condition alternative to the one used by Renes
\emph{et al}~\cite{Renes}.
Renes \emph{et al}~\cite{Renes} base their numerical construction of
Weyl-Heisenberg covariant SI(1)-POVMs on the fact that, if $\hat{P}$ is an
arbitrary  rank 1 projector and $\hat{P}_{\mathbf{p}} =
\hat{D}^{\vphantom{\dagger}}_{\mathbf{p}}\hat{P}
\hat{D}^{\dagger}_{\mathbf{p}}$, then
\begin{equation}
\sum_{\mathbf{p} \in \mathbb{Z}_d^2} \bigl(\Tr (\hat{P}_{\mathbf{p}}
\hat{P})\bigr)^2 \ge \frac{2 d}{d+1}
\label{eq:RBSCExtremCond}
\end{equation}
with equality if and only if the   operators $\frac{1}{d}
\hat{P}_{\mathbf{p}}$ constitute an SI(1)-POVM.  The inequality we have
derived provides us with an alternative procedure:  instead of looking for
a projector $\hat{P}$ which minimizes the expression on the left hand side
of Eq.~(\ref{eq:RBSCExtremCond}), one can look for a set of phases which
maximize the expression on the left hand side of
Eq.~(\ref{eq:PhaseExtremCond}).

It should be said that if one is specifically looking for a method of
constructing SI(1)-POVMs numerically a procedure based on
Eq.~(\ref{eq:RBSCExtremCond}) is likely to be more efficient than  one
based on Eq.~(\ref{eq:PhaseExtremCond}).  This is because the expression
on the left hand side of Eq.~(\ref{eq:RBSCExtremCond}) is a function of
$2(d-1)$ real parameters (\emph{i.e.}\ the number of parameters needed to
specify the projector $\hat{P}$), whereas the one on the left hand side of
Eq.~(\ref{eq:PhaseExtremCond}) is a function of
$(d^2+1)/2$ real parameters if
$d$ is odd and
$(d^2+4)/2$ real parameters if $d$ is even (\emph{i.e.}\ the number of
independent phase angles).

However, although the extremal condition represented by
Eq.~(\ref{eq:PhaseExtremCond}) would appear not to have any advantages
from a concrete numerical point of view, it may perhaps be interesting
from a more abstract mathematical point of view, as providing additional
insight into the problem.  In particular, the fact that the phase angles
appear in combinations of the form
$\theta_{\mathbf{p}}+\theta_{\mathbf{q}}+\theta_{\mathbf{r}}$ with
$\mathbf{p}+\mathbf{q}+\mathbf{r}=\boldsymbol{0}\;\text{(mod $d$)}$ may
possibly provide some clues as to the origin of the order
$3$ symmetry found in every Weyl-Heisenberg covariant SI(1)-POVM constructed
to  date\footnote{Note that Grassl~\cite{GrasslB} has constructed a
counter-example in dimension $12$ to conjecture C of
ref.~\cite{ApplebyB}.  However, his example is still invariant under a
canonical order $3$ unitary.  Specifically his matrix
$T_{12}$ is a representative of the Clifford operation
$\biggl[\begin{pmatrix} 4 & 3 \\ 9 &7\end{pmatrix},
\begin{pmatrix} -3 \\ -6 \end{pmatrix}
\biggr]$, which it will be seen has Clifford trace $=-1$ (notation and
terminology as in ref.~\cite{ApplebyB}).  His example is therefore
consistent with conjecture A of ref.~\cite{ApplebyB}. }.

\section{The Wigner POVM}
\label{sec:WigPOVM}
Suppose that $d$ is odd.  In that case we can set the phase angles on the
right hand side of Eq.~(\ref{eq:WHGeneratorTermsPhases}) equal to zero,
giving
\begin{equation}
\hat{B} = \frac{1}{\sqrt{d+1}} \sum_{\mathbf{q}\in(\mathbb{Z}_d^2)^{*}}
\hat{D}_{\mathbf{q}}
\end{equation}
(notice  that if $d$ was even this choice of  phases   would not be
permissible because when $d$ is even the signs $s_{-\mathbf{q}}$ are not
all positive). For reasons explained below we will refer to the
SI-POVM corresponding to this choice of $\hat{B}$  as the Wigner POVM. 

We wish to determine the efficiency parameter of  the Wigner POVM.  For
that purpose it is convenient to consider the operator\footnote{In the
notation of ref.~\cite{ApplebyB} $\hat{U}$ is a representative of the
Clifford operation $\left[\begin{pmatrix} -1 & 0 \\ 0 & -1\end{pmatrix},
\begin{pmatrix} 0 \\ 0
\end{pmatrix}\right]$.  Its action on the standard basis used to define
the operators $\hat{D}_{\mathbf{p}}$ (see Eqs.~(\ref{eq:TDef})
and~(\ref{eq:SDef})) is $\hat{U}|r\rangle = |\bar{r}\rangle $.  So
$\hat{U}$ can be thought of as  a discrete parity operator.
 }
\begin{equation}
\hat{U} = \frac{1}{d} (1+\sqrt{d+1} \hat{B}) = \frac{1}{d}
\sum_{\mathbf{q}\in\mathbb{Z}_d^2} \hat{D}_{\mathbf{q}}
\label{eq:ParityOpDef}
\end{equation}
$\hat{U}$, like $\hat{B}$, is an  Hermitian operator.   Moreover
\begin{align}
\hat{U}^2 & =\frac{1}{d^2}
\sum_{\mathbf{q},\mathbf{r}\in\mathbb{Z}_d^2}\tau^{\langle
\mathbf{q},\mathbf{r}\rangle}\hat{D}_{\mathbf{q}+\mathbf{r}}
\nonumber
\\
& =\frac{1}{d^2} \sum_{\mathbf{q},\mathbf{r}\in\mathbb{Z}_d^2}
\tau^{\langle
\mathbf{q},\mathbf{r}\rangle} \hat{D}_{\mathbf{q}}
\nonumber
\\
& = 1
\end{align} 
where we used the fact that
 $\sum_{\mathbf{r}\in\mathbb{Z}_d^2}\tau^{\langle
\mathbf{q},\mathbf{r}\rangle}  = d^2 \delta_{\mathbf{q}\boldsymbol{0}}$
(note that this depends on the fact that $d$ is odd).  It follows that the
eigenvalues of $\hat{U}$ all $=\pm 1$.  Taking into account the fact that 
$\Tr(\hat{U})=1$ we deduce that $\hat{U}$ must have $(d+1)/2$ eigenvalues
$=1$ and $(d-1)/2$ eigenvalues $=-1$.  Consequently the smallest
eigenvalue of
$\hat{B}$ is
$-\sqrt{d+1}$.  In view of Theorem~\ref{thm:EvalThm2} it follows that
$(1/\sqrt{d+1}) \hat{B}$ is a Bloch vector.  Hence the $d^2$ operators
\begin{equation}
\hat{E}_{\mathbf{p}} = \frac{1}{d^2}
\left(1+\frac{1}{\sqrt{d+1}}\hat{B}_{\mathrm{p}}
\right)
\label{eq:WigPOVMdef}
\end{equation}
constitute an SI-POVM of rank $(d+1)/2$.  We will refer to this as the
Wigner POVM.  It has efficiency parameter $1/\sqrt{d+1}$---which is
a considerable improvement on the worst case value $1/(d-1)$ calculated in
Section~\ref{sec:WHCovariantSIPOVMs}, although still greatly inferior to
the best case value $\kappa=1$.

Let us now explain the connection between the Wigner POVM and the Wigner
function.  Let $\hat{\rho}$ be an arbitrary density matrix, and let
\begin{equation}
\rho_{\mathbf{p}} = \frac{1}{d} \Tr(\hat{D}^{\dagger}_{\mathbf{p}}
\hat{\rho} )
\label{eq:rhopDef}
\end{equation}
  We
define the
 Wigner function $W_{\mathbf{p}}$ to be the discrete Fourier transform of
the
 coefficients $\rho_{\mathbf{p}}$:
\begin{equation}
W_{\mathbf{p}} = \frac{1}{d} \sum_{\mathbf{q}\in\mathbb{Z}_d^2} \tau^{-2
\langle \mathbf{p}, \mathbf{q} \rangle} \hat{\rho}_{\mathbf{p}}
\label{eq:WigDef}
\end{equation}
This definition agrees with that of
Wootters~\cite{WoottersB} in the case when $d$ is prime.  If $d$ is
non-prime the  Wigner function as defined by this formula loses some of
the  properties which Wootters considers desirable.  However, it appears
to us that it retains sufficiently many of these properties for it still
to be considered a reasonable way of defining the Wigner function.  For
further discussion of the discrete Wigner function see
refs.~\cite{WoottersB,WoottersE,Miquel,Vourdas,Chat,Gross} and references
cited therein.

The Wigner function can   be expressed in terms of the
operators
$\hat{U}_{\mathbf{p}} = 
\hat{D}^{\vphantom{\dagger}}_{\mathbf{p}} \hat{U}
\hat{D}^{\dagger}_{\mathbf{p}}$ (where $\hat{U}$ is 
the operator defined in
Eq.~(\ref{eq:ParityOpDef})).  In fact
\begin{equation}
\hat{U}_{\mathbf{p}}
=\frac{1}{d} \sum_{\mathbf{q}\in\mathbb{Z}_d^2} \tau^{2
\langle \mathbf{p}, \mathbf{q} \rangle} \hat{D}_{\mathbf{q}}
=\frac{1}{d} \sum_{\mathbf{q}\in\mathbb{Z}_d^2} \tau^{-2
\langle \mathbf{p}, \mathbf{q} \rangle} \hat{D}^{\dagger}_{\mathbf{q}}
\end{equation}
Eqs.~(\ref{eq:rhopDef}) and~(\ref{eq:WigDef}) then imply
\begin{equation}
W_{\mathbf{p}} = \frac{1}{d} \Tr( \hat{U}_{\mathbf{p}}\hat{\rho})
\end{equation}
Taking into account Eqs.~(\ref{eq:ParityOpDef}) and~(\ref{eq:WigPOVMdef})
we deduce
\begin{equation}
W_{\mathbf{p}} = (d+1) \Tr(\hat{E}_{\mathbf{p}} 
\hat{\rho} ) - \frac{1}{d}
\label{eq:WigFnTermsWigPOVM}
\end{equation}
Of course, the fact that the Wigner function is a linear function of the
probabilities $\Tr(\hat{E}_{\mathbf{p}} \hat{\rho})$ is an automatic
consequence of the fact that the POVM is informationally complete. 
However, in the case of the Wigner POVM the relationship is particularly
simple:  to obtain the Wigner function one merely has to rescale the
probabilities by a constant amount and then  shift them by a constant
amount.

 Eq.~(\ref{eq:WigFnTermsWigPOVM}) is conceptually interesting because it
establishes a connection between  SI-POVMs and the  Wigner function.  At
first sight it may appear that it also has a more concrete, pragmatic
significance, as providing  a good way to determine the Wigner function
tomographically.   However, a little reflection will dispel
that impression.  The trouble is that the Wigner POVM has efficiency
parameter $=1/\sqrt{d+1}$, which is $<1$ (and $\ll 1$ if $d$ is large). 
So if one wants to determine the numbers $W_{\mathbf{p}}$ it would be
much more efficient (would give much less statistical uncertainty for a
given number of measurements) to use a tomographic scheme based on an
SI(1)-POVM, or a full set of MUBs (in dimensions where such exist), and
then to perform the appropriate linear transformation on the relative
frequencies obtained~\cite{ApplebyA}.

Finally, let us note that Miquel \emph{et al}~\cite{Miquel} have
described a  scheme for ``directly measuring'' the individual numbers
$W_{\mathbf{p}}$.  This
scheme might, perhaps, have some advantages over a scheme based on an
SI(1)-POVM or  a full set of MUBs in a case where one was only interested
in
\emph{some} of the numbers  
$W_{\mathbf{p}}$.

\section{Conclusion}
We originally undertook the investigations reported here in the hope that
they might lead to a solution of the really challenging problem, which is
to demonstrate the existence (or, as it may be, the non-existence) of
SI(1)-POVMs in every finite dimension.  We did not succeed in that primary
aim.  Nevertheless, we derive some consolation from the fact that the
class of SI-POVMs is of some intrinsic interest.  Also, it is not
impossible that the results reported here contain  clues that may help us
to solve the main problem.

\appendix
\section{Weyl-Heisenberg Group}
\label{sec:WHGroup}
In this appendix we summarise those facts concerning the Weyl-Heisenberg
group (or generalized Pauli group as it is sometimes called) which are
needed in the main text. Our definitions are those of ref.~\cite{ApplebyB},
and may differ slightly from the ones used by other authors.   Let
$|0\rangle, |1\rangle, \dots |d-1\rangle$ be an orthonormal basis for
$\mathcal{H}$, and 
let\footnote{The reason for defining $\tau = - e^{\pi i /d}$ rather than 
$\tau = e^{\pi i /d}$ is that it means $\tau^{d^2} =1$ for all $d$. 
 }
$\tau = - e^{\pi i /d}$.  Define operators
$\hat{T}$ and $\hat{S}$ by
\begin{align}
 \hat{T} |r\rangle & =\tau^{2 r} |r\rangle
\label{eq:TDef}
\\
  \hat{S} |r\rangle &  = \begin{cases} 
  |r+1 \rangle \qquad & r=0,1,\dots , d-2
\\
  |0\rangle  \qquad & r=d-1
\end{cases}
\label{eq:SDef} 
\\
\intertext{ 
Then define, for each pair of integers $\mathbf{p}=(p_1,p_2)\in
\mathbb{Z}^2$,} 
  \hat{D}_{\mathbf{p}} & =\tau^{p_1 p_2} \hat{S}^{p_1} \hat{T}^{p_2}
\label{eq:DOpDefinition}
\end{align}
The operators $\hat{D}_{\mathbf{p}}$ are the displacement operators of
the Weyl-Heisenberg group.  The reason for including the factor
$\tau^{p_1 p_2}$ is that it means that the operators have the following
nice properties:
\begin{align}
 \hat{D}_{\mathbf{p}}^\dagger & = \hat{D}_{-\mathbf{p}} 
\label{eq:DConjugate}
\\
  \hat{D}_{\mathbf{p}} \hat{D}_{\mathbf{q}} & = \tau^{\langle
\mathbf{p},\mathbf{q}\rangle} \hat{D}_{\mathbf{p} + \mathbf{q}}
\label{eq:DCompositionRule}
\intertext{where}
\langle
 \mathbf{p},\mathbf{q}\rangle & =p_2 q_1-p_1 q_2
\end{align} The fact that $\langle
\mathbf{p},\mathbf{p}\rangle=0$ means
\begin{equation}
\bigl(\hat{D}_{\mathbf{p}}\bigr)^n = \hat{D}_{n \mathbf{p}}
\end{equation}
for all $\mathbf{p} \in \mathbb{Z}^2$ and $n\in \mathbb{Z}$. 
In particular the operators $\hat{D}_{\mathbf{p}}$ are unitary:
\begin{equation}
\hat{D}^{\dagger}_{\mathbf{p}}
\hat{D}^{\vphantom{\dagger}}_{\mathbf{p}} = 1
\end{equation}
for all $\mathbf{p}$.  It is also worth noting that
\begin{equation}
\bigl(\hat{D}_{\mathbf{p}}\bigr)^d = 1
\end{equation}
for all $\mathbf{p}$ (this is one of the reasons for setting $\tau = -
e^{i \pi/d}$.  If, instead, one set $\tau = 
e^{i \pi/d}$ it would sometimes happen that
$\bigl(\hat{D}_{\mathbf{p}}\bigr)^d = -1$).

The presence of the factor $\tau^{\langle
\mathbf{p},\mathbf{q}\rangle}$ on the right hand side of
Eq.~(\ref{eq:DCompositionRule}) means that the operators
$\hat{D}_{\mathbf{p}}$ do not constitute a group.  However, one obtains a
group (the Weyl Heisenberg group) if one takes the set of all operators of
the form
$e^{ i\alpha}
\hat{D}_{\mathbf{p}}$, where $e^{i \alpha}$ is an arbitrary phase
(alternatively, one can define the Weyl-Heisenberg group to be the set of
all operators of the form $\tau^n
\hat{D}_{\mathbf{p}}$, where $n$ is an arbitrary integer).

If $\mathbf{p}=\mathbf{q}  \text{ (mod } d\text{)}$ then
$\hat{D}_{\mathbf{p}} =
\hat{D}_{\mathbf{q}}$ up to a sign.   Specifically:
\begin{equation}
\hat{D}_{\mathbf{p}}
=
\begin{cases} \hat{D}_{\mathbf{q}} \quad & \text{if $d$ is odd} \\
(-1)^{\frac{1}{d} \langle
\mathbf{p},\mathbf{q}\rangle}\hat{D}_{\mathbf{q}}
\quad &
\text{if
$d$ is even}
\end{cases}
\label{eq:DOpModdRel}
\end{equation}
(to prove this formula  write
$\mathbf{p} =\mathbf{q} + d\mathbf{u}$ and then use
Eq.~(\ref{eq:DCompositionRule})).  It is therefore often convenient to
restrict ourselves to values of $\mathbf{p}$ lying in the set
$\mathbb{Z}_d^2=\{(p_1,p_2)\colon p_1,p_2 =0,1,2,\dots , d-1\}$. Given
arbitrary $\mathbf{p} \in\mathbb{Z}^2$ let $[\mathbf{p}]$ be the unique
element of $\mathbb{Z}_{d}^{2}$ such that $[\mathbf{p}] = \mathbf{p} \;
\text{mod $d$}$.  It is also convenient to define
\begin{align}
\mathbf{p} \oplus \mathbf{q} & = [\mathbf{p} + \mathbf{q}]
\\
\mathbf{p} \ominus \mathbf{q} & = [\mathbf{p} - \mathbf{q}]
\\
\bar{\mathbf{p}} & = [-\mathbf{p}]
\end{align}
and
\begin{equation}
s_{\mathbf{p}}  = \begin{cases} 1 \quad &  \text{if $d$ is odd}\\
(-1)^{\frac{1}{d} \langle \mathbf{p},[\mathbf{p}]\rangle}
\quad & \text{if $d$ is even}
\end{cases}
\end{equation}
We then have, for all $\mathbf{p}, \mathbf{q} \in \mathbb{Z}_d^2$,
\begin{align}
 \hat{D}_{\mathbf{p}}^\dagger & = s_{-\mathbf{p}}
\hat{D}_{\bar{\mathbf{p}}} 
\label{eq:DdagTermsBarp}
\\
  \hat{D}_{\mathbf{p}} \hat{D}_{\mathbf{q}} & =
s_{\mathbf{p}+\mathbf{q}} \tau^{\langle
\mathbf{p},\mathbf{q}\rangle} \hat{D}_{\mathbf{p} \oplus \mathbf{q}}
\end{align}
It is also easily verified that 
\begin{equation}
\Tr \bigl(\hat{D}^{\dagger}_{\mathbf{p}}
\hat{D}^{\vphantom{\dagger}}_{\mathbf{q}}
\bigr)
= d \delta_{\mathbf{p} \mathbf{q}}
\end{equation}
for all $\mathbf{p}$, $\mathbf{q}\in \mathbb{Z}_d^2$.  This means that,
relative to the Hilbert-Schmidt inner product, the
operators $\frac{1}{\sqrt{d}}\hat{D}_{\mathbf{p}} $ are an orthonormal
basis for the $d^2$ complex dimensional space $\mathcal{L}(\mathcal{H})$
consisting of all
$d\times d$ complex matrices. So an arbitrary matrix
$\hat{A}\in\mathcal{L}(\mathcal{H})$ can be expanded
\begin{equation}
\hat{A} = \sum_{\mathbf{p} \in \mathbb{Z}_d^2} A_{\mathbf{p}}
\hat{D}_{\mathbf{p}}
\label{eq:ArbOpTermsDExp}
\end{equation}
where the expansion coefficients are given by
\begin{equation}
A_{\mathbf{p}} = \frac{1}{d} \Tr \bigl( \hat{D}^{\dagger}_{\mathbf{p}}
\hat{A}
\bigr)
\label{eq:WHExpCoeffs}
\end{equation}
It follows from Eqs.~(\ref{eq:DdagTermsBarp}), (\ref{eq:ArbOpTermsDExp})
and~(\ref{eq:WHExpCoeffs}) that  $\hat{A}$ is Hermitian if and
only if
\begin{equation}
A^{\vphantom{*}}_{\bar{\mathbf{p}}} = 
s^{\vphantom{*}}_{-\mathbf{p}} A^{*}_{\mathbf{p}}
\label{eq:HermCond}
\end{equation}
for all $\mathbf{p}\in \mathbb{Z}_d^2$.

Let $(\mathbb{Z}^2_d)^{*}=\{\mathbf{p}\in\mathbb{Z}^2_d\colon \mathbf{p}
\neq (0,0) \}$.  The fact that 
\begin{equation}
\Tr (\hat{D}_{\mathbf{p}}) = \begin{cases} d \quad & \text{if $\mathbf{p}
= \boldsymbol{0}$ (mod $d$)}
\\ 0 \quad & \text{otherwise}
\end{cases}
\label{eq:TrDExp}
\end{equation}
 means that $\hat{A} \in
\mathrm{su}(d)$ if and only if it has an expansion
\begin{equation}
\hat{A} = \sum_{\mathbf{p} \in (\mathbb{Z}_d^2)^{*}}
A_{\mathbf{p}}
\hat{D}_{\mathbf{p}}
\label{eq:sudExpansion}
\end{equation}
where the coefficients satisfy Eq.~(\ref{eq:HermCond}).

The following lemma tells us how to calculate the expansion coefficients
and traces of double and triple products:
\begin{lemma}
\label{thm:TraceFormulae}
Let $\hat{A}, \hat{B}, \hat{C}\in
\mathcal{L}(\mathcal{H})$. Then
\begin{align}
\bigl( \hat{A} \hat{B} \bigr)_{\mathbf{p}}
& =
\sum_{\mathbf{q}\in\mathbb{Z}_d^2} 
s_{\mathbf{p} - \mathbf{q}} \tau^{\langle \mathbf{q},\mathbf{p}\rangle}
A_{\mathbf{q}}B_{\mathbf{p}\ominus\mathbf{q}}
\label{eq:DoubleProductExp}
\\
\bigl( \hat{A} \hat{B} \hat{C} \bigr)_{\mathbf{p}}
& =
\sum_{\mathbf{q},\mathbf{r} \in \mathbb{Z}^2_d}
s_{\mathbf{p}-\mathbf{q}-\mathbf{r}}
\tau^{ \langle\mathbf{q}+
\mathbf{r}, \mathbf{p}\rangle+\langle \mathbf{q}, \mathbf{r} \rangle }
A_{\mathbf{q}} B_{\mathbf{r}}
C_{\mathbf{p}\ominus\mathbf{q}\ominus\mathbf{r}}
\label{eq:TripleProductExp}
\end{align}
where 
$(\hat{A}\hat{B})_{\mathbf{p}} =(1/d)
\Tr(\hat{D}^{\dagger}_{\mathbf{p}} \hat{A}\hat{B})$ and  
$(\hat{A}\hat{B}\hat{C})_{\mathbf{p}}=(1/d)\Tr(\hat{D}^{\dagger}_{\mathbf{p}}
\hat{A}\hat{B} \hat{C}) $ are the expansion coefficients as given by
Eq.~(\ref{eq:WHExpCoeffs}).  Traces are given by
\begin{align}
\Tr(\hat{A}\hat{B}) & = d \sum_{\mathbf{q}\in\mathbb{Z}_2^d}
s_{-\mathbf{q}} A_{\mathbf{q}} B_{\bar{\mathbf{q}}}
\label{eq:DoubleTraceExp}
\\
\Tr(\hat{A}\hat{B}\hat{C}) & = d \sum_{\mathbf{q},
\mathbf{r}\in\mathbb{Z}_2^d}
s_{-\mathbf{q}-\mathbf{r}}
\tau^{\langle \mathbf{q},\mathbf{r}\rangle} A_{\mathbf{q}}
B_{\mathbf{r}} C_{\bar{\mathbf{q}}\oplus\bar{\mathbf{r}}}
\label{eq:TripleTraceExp}
\end{align}
If $\hat{A}, \hat{B}, \hat{C}$  are  Hermitian we can alternatively
write
\begin{align}
\Tr(\hat{A}\hat{B}) & = d \sum_{\mathbf{q}\in\mathbb{Z}_2^d}
A^{\vphantom{*}}_{\mathbf{q}} B^{*}_{\mathbf{q}}
\label{eq:HermDoubleTraceExp}
\\
\Tr(\hat{A}\hat{B}\hat{C}) & = d \sum_{\mathbf{q},
\mathbf{r}\in\mathbb{Z}_2^d} s_{\mathbf{q}+\mathbf{r}}
\tau^{\langle \mathbf{q},\mathbf{r}\rangle} A^{\vphantom{*}}_{\mathbf{q}}
B^{\vphantom{*}}_{\mathbf{r}} C^{*}_{\mathbf{q}\oplus\mathbf{r}}
\label{eq:HermTripleTraceExp}
\end{align}
\end{lemma}
\begin{proof}
Let $\hat{A}, \hat{B}\in\mathcal{L}(\mathcal{H})$.  It follows from
Eq.~(\ref{eq:ArbOpTermsDExp}) that
\begin{align}
\hat{A}\hat{B} & = \sum_{\mathbf{q},\mathbf{r}\in\mathbb{Z}^2_d}
A_{\mathbf{q}}B_{\mathbf{r}} \hat{D}_{\mathbf{q}}\hat{D}_{\mathbf{r}}
\nonumber
\\
& = \sum_{\mathbf{q},\mathbf{r}\in\mathbb{Z}^2_d}
s_{\mathbf{q}+\mathbf{r}}
\tau^{\langle\mathbf{q},\mathbf{r}\rangle}
A_{\mathbf{q}}B_{\mathbf{r}}
\hat{D}_{\mathbf{q}\oplus\mathbf{r}}
\nonumber
\\
&  =\sum_{\mathbf{q},\mathbf{p}\in\mathbb{Z}^2_d}
s_{\mathbf{q}+\mathbf{p}\ominus\mathbf{q}}
\tau^{\langle\mathbf{q},\mathbf{p}\ominus\mathbf{q}\rangle}
A_{\mathbf{q}}B_{\mathbf{p}\ominus\mathbf{q}}
\hat{D}_{\mathbf{p}}
\nonumber
\\
& = \sum_{\mathbf{p}\in\mathbb{Z}^2_d}
\Biggl(\sum_{\mathbf{q}\in\mathbb{Z}^2_d}s_{\mathbf{p}-\mathbf{q}}
\tau^{\langle \mathbf{q},\mathbf{p}\rangle}
A_{\mathbf{q}}B_{\mathbf{p}\ominus\mathbf{q}}
\Biggr) 
\hat{D}_{\mathbf{p}}
\end{align}
where in the last line we used the fact that $
s_{\mathbf{q}+\mathbf{p}\ominus\mathbf{q}}
\tau^{\langle\mathbf{q},\mathbf{p}\ominus\mathbf{q}\rangle}
=
s_{\mathbf{p}-\mathbf{q}} \tau^{\langle \mathbf{q},\mathbf{p}\rangle}$.
Eq.~(\ref{eq:DoubleProductExp}) is now immediate.

To prove Eq.~(\ref{eq:TripleProductExp})  we apply
Eq.~(\ref{eq:DoubleProductExp}) twice:
\begin{align}
(\hat{A}\hat{B}\hat{C})_{\mathbf{p}}
& = \sum_{\mathbf{q}\in\mathbb{Z}_d^2}
s_{\mathbf{p}-\mathbf{q}}\tau^{\langle\mathbf{q},\mathbf{p}\rangle}
A_{\mathbf{q}} ( \hat{B}\hat{C})_{\mathbf{p}\ominus\mathbf{q}}
\nonumber
\\
& =
\sum_{\mathbf{q},\mathbf{r}\in\mathbb{Z}_d^2}
s_{\mathbf{p}-\mathbf{q}}s_{\mathbf{p}\ominus\mathbf{q}-\mathbf{r}}
\tau^{\langle\mathbf{q},\mathbf{p}\rangle+\langle
\mathbf{r},\mathbf{p}\ominus\mathbf{q}\rangle} A_{\mathbf{q}}
B_{\mathbf{r}} C_{\mathbf{p}\ominus\mathbf{q}\ominus\mathbf{r}}
\nonumber
\\
& = 
\sum_{\mathbf{q},\mathbf{r}\in\mathbb{Z}_d^2}
s_{\mathbf{p}-\mathbf{q}-\mathbf{r}}
\tau^{\langle\mathbf{q}+\mathbf{r},\mathbf{p}\rangle+\langle
\mathbf{q},\mathbf{r}\rangle} A_{\mathbf{q}}
B_{\mathbf{r}} C_{\mathbf{p}\ominus\mathbf{q}\ominus\mathbf{r}}
\end{align}
where in the last line we used the identity
$s_{\mathbf{p}-\mathbf{q}}s_{\mathbf{p}\ominus\mathbf{q}-\mathbf{r}}
\tau^{\langle
\mathbf{r},\mathbf{p}\ominus\mathbf{q}\rangle}=
s_{\mathbf{p}-\mathbf{q}-\mathbf{r}}\tau^{\langle
\mathbf{r},\mathbf{p}-\mathbf{q}\rangle}$.

To prove Eqs.~(\ref{eq:DoubleTraceExp}) and~(\ref{eq:TripleTraceExp}) 
set $\mathbf{p}=\boldsymbol{0}$ in Eqs.~(\ref{eq:DoubleProductExp})
and~(\ref{eq:TripleProductExp}) and use the fact that
$\Tr(\hat{M}) = d M_{\boldsymbol{0}}$  for all
$\hat{M}\in\mathcal{L}(\mathcal{H})$. Eq.~(\ref{eq:HermDoubleTraceExp})
follows from Eq.~(\ref{eq:DoubleTraceExp}). 
Eq.~(\ref{eq:HermTripleTraceExp}) follows from
Eq.~~(\ref{eq:TripleTraceExp}) and the identity
$s_{\mathbf{p}+\mathbf{q}} s_{-\mathbf{p}-\mathbf{q}} =
s_{-\mathbf{p}\oplus\mathbf{q}}$.
\end{proof}


\begin{thebibliography}{99}
 \bibitem{Renes}  J.M.~Renes, R.~Blume-Kohout, A.J.~Scott  and
C.M.~Caves,
\emph{J.\ Math.\ Phys.}\ \textbf{45}, 2171 (2004).  Also available
as quant-ph/0310075.
\bibitem{FuchsA} C.A.~Fuchs, \emph{Quantum Information
and Computation} \textbf{4}, 467 (2004). Also available
as quant-ph/0404122.
\bibitem{WoottersC} W.K.~Wootters, quant-ph/0406032.
\bibitem{BengA}  I.~Bengtsson, quant-ph/0406174.
\bibitem{GrasslA}  M.~Grassl, in  \emph{Proceedings ERATO
Conference on Quantum Information Science 2004} (Tokyo,
2004).   Also available as quant-ph/0406175.
\bibitem{BengB} I.~Bengtsson and A.~Ericsson, 
\emph{Open Sys.\ and Information Dyn.}\ \textbf{12}, 187 (2005).  Also
available as quant-ph/0410120.
\bibitem{ApplebyB} D.M.~Appleby,\emph{J.\ Math.\ Phys.}\ \textbf{46},
052107 (2005).  Also available as quant-ph/0412001.
\bibitem{GrasslB} M.~Grassl, \emph{Electronic Notes in Discrete
Mathematics} \textbf{20}, 151 (2005).
\bibitem{KlappA} A.~Klappenecker and M.~R\"{o}tteler, quant-ph/0502031.
\bibitem{KlappB} A.~Klappenecker, M.~R\"{o}tteler, I.~Shparlinski and
A.~Winterhof, quant-ph/0503239.
\bibitem{Scott} A.J.~Scott, quant-ph/0604049.
\bibitem{Flammia} S.T.~Flammia, quant-ph/0605050.
\bibitem{Kim} I.H.~Kim, quant-ph/0608024.
\bibitem {Prugo}  E.~Prugove\v{c}ki,  \emph{Int.\ J.\ Theor.\ Phys.}\
\textbf{16}, 321 (1977).
\bibitem{Schroeck} F.E.~Schroeck,  \emph{Int.\ J.\ Theor.\ Phys.}\
\textbf{28}, 247 (1989).
\bibitem{BuschA}  P.~Busch, \emph{Int.\ J.\ Theor.\ Phys.}\
\textbf{30}, 121 (1991).
\bibitem{BuschB} P.~Busch, M.~Grabowski and P.J.~Lahti, \emph{Operational
Quantum Physics} (Springer, Lecture Notes in Physics m31, 1995).
\bibitem{Caves} C.M.~Caves, C.A.~Fuchs and R.~Schack, 
\emph{J.\ Math.\ Phys.}\ \textbf{43}, 4537 (2002). Also available as
quant-ph/0104088.
\bibitem{FuchsB} C.A.~Fuchs, quant-ph/0205039. 
\bibitem{dAriano} G.M.~d'Ariano, P.~Perinotti  and M.F.~Sacchi,,
\emph{J.\ Opt.\ B:  Quantum and Semicl.\ Optics}, \textbf{6}, S487
(2004).  Also available as quant-ph/0310013.
\bibitem{Hoggar} S.G.~Hoggar, 
\emph{Geom.\ Dedic.}\ \textbf{69} (1998).
\bibitem{Zauner} G.~Zauner, ``Quantum designs---foundations of a
non-commutative theory of designs'' (in German), Ph.D. thesis, University
of Vienna, 1999.  Available online at
http://www.mat.univie.ac.at/\~{}neum/papers/physpapers.html.
\bibitem{Harriman} J.E.~Harriman, \emph{Phys.\ Rev.\ A} \textbf{17},
1249 (1978).
\bibitem{Mahler} G.~Mahler and V.A.~Weberruss, \emph{Quantum Networks: 
Dynamics of Open Nanostructures} (Springer, Berlin, 1995).
\bibitem{Jakobczyk} L.~Jak\'{o}bczyk and M.~Siennicki, \emph{Phys.\
Lett.\ A} \textbf{286}, 383 (2001).
\bibitem{Kim1} G.~Kimura, \emph{Phys.\ Lett.\ A}
\textbf{314}, 339 (2003).  Also available as quant-ph/0301152.
\bibitem{Byrd} M.S.~Byrd and N.~Khaneja, \emph{Phys.\ Rev.\ A}
\textbf{68}, 062322 (2003). Also available as quant-ph/0302024.
\bibitem{Schirmer} S.G.~Schirmer, T.~Zhang and J.V.~Leahy, 
\emph{J.\ Phys.\ A} \textbf{37}, 1389 (2004).  Also available as
quant-ph/0308004.
\bibitem{Kim2} G.~Kimura and A.~Kossakowski, \emph{Open Sys.\ Information
Dyn.}\ \textbf{12}, 207 (2005).  Also available as quant-ph/0408014.
\bibitem{Dietz} K.~Dietz, quant-ph/0601013.
\bibitem{Ivanovic}  I.D.~Ivanovi\'{c},  \emph{J. Phys. A} \textbf{14},
3241 (1981).
\bibitem{WoottersB} W.K.~Wootters,  \emph{Ann. Phys.
(N.Y.)} \textbf{176}, 1 (1987).
\bibitem{WoottersA} W.K.~Wootters and B.D.~Fields,  \emph{Ann. Phys.
(N.Y.)} \textbf{191}, 363 (1989).
\bibitem{Bandy} S.~Bandyopadhyay, P.O.~Boykin, V.~Roychowdhury and
F.~Vatan, quant-ph/0103162.
\bibitem{Pitt}  A.O. Pittenger and M.H. Rubin, 
\emph{Linear Alg.\ Appl.}\ \textbf{390}, 255 (2004).  Also available as
quant-ph/0308142.
\bibitem{KlappC}  A.~Klappenecker and M.~R\"{o}tteler, 
quant-ph/0309120.
\bibitem{Archer} C.~Archer, quant-ph/0312204.
\bibitem{Durt} T.~Durt,  quant-ph/0401046.
\bibitem{WoottersE}  K.S.~Gibbons, M.J.~Hoffman, W.K.~Wootters
\emph{Phys. Rev. A} \textbf{70}, 062101 (2004).  Also available as
quant-ph/0401155.
\bibitem{Planat} M.~Saniga, M.~Planat and H.~Rosu, \emph{J.\ Opt.\ B: 
Quantum and Semicl.\ Optics}, \textbf{6},  L19 (2004).  Also
available as math-ph/0403057.
\bibitem{ApplebyA}  D.M.~Appleby, to appear.
\bibitem{Miquel} C.~Miquel, J.P.~Paz and M.~Saraceno, 
\emph{Phys.\ Rev.\ A} \textbf{65}, 062309 (2002).  Also available as
quant-ph/0204149.
\bibitem{Vourdas} A.~Vourdas, \emph{Rep.\ Prog.\ Phys.}\ \textbf{67},
267 (2004).
\bibitem{Chat} S.~Chaturvedi, E.~Ercolessi, G.~Marmo, G.~Morandi,
N.~Mukunda, R.~Simon, \emph{Pramana} \textbf{65}, 981 (2006).  Also
available as quant-ph/0507094.
\bibitem{Gross} D.~Gross, quant-ph/0602001.
\end{thebibliography}
\end{document}